# Review on Giant Magnetoelectric effects in Oxide ferromagnetic/ferroelectric Layered Structures

*G. Srinivasan*

*Physics Department, Oakland University, Rochester, MI 4830, USA*

*Abstract*

*The synthesis of layered ferrite-lead titanate zirconate (PZT) and lanthanum nanganite-PZT and the observation of giant magneto-electric interactions are discussed. The ferrites used in our studies included pure and Zn substituted cobalt-, nickel- and lithium ferrites. Ferromagnetic manganites used included both calcium and strontium substituted systems. The samples were prepared from 10-40 μm thick films obtained by tape-casting. Our studies showed strong ME coupling in manganite-PZT and cobalt zinc ferrite-PZT, and a giant ME effect in nickel zinc ferrite-PZT. We found evidence for Zn assisted enhancement in ME coefficients that were attributed to low anisotropy and high permeability that resulted in favorable magneto-mechanical coupling in the composites.*

## 1. Introduction

In materials that are magnetoelectric, the induced polarization P is related to the field H by the expression, $P = \alpha H$, where $\alpha$ is the second rank ME-susceptibility tensor and is expressed in the units of s/m in SI units [in Gaussian units $\alpha = 4\pi P/H = 4\pi M/E$ is dimensionless]. One generally determines $\alpha$ by measuring $\delta P$ for an applied ac field $\delta H$. Another parameter of importance is the ME voltage coefficient $\alpha_E = \delta E/\delta H$ which is related to $\alpha$ by the expression $\alpha = \varepsilon_o \varepsilon_r \alpha_E$, where $\varepsilon_r$ is the relative permitivity of the material.

In order to improve material characteristics or for the engineering of materials with desired properties Van Suchtelen proposed two types of composites, sum properties and product properties [1]. A product-property composite consisting of piezomagnetic-piezoelectric phases is expected to be magnetoelectric since the ME coefficient $\alpha = \delta P/\delta H$ is the product of the piezomagnetic deformation $\delta z/\delta H$ and the piezoelectric charge generation $\delta Q/\delta z$. The ME composite of primary interest in the past was $CoFe_2O_4$-$BaTiO_3$. Although cobalt ferrite is not piezomagnetic, lattice deformation due to high magnetostriction gives rise to ME effects in the mixed oxide. For this specific composite under ideal conditions, Van den Boomgard [2] estimated an upper limit for the coefficient $\alpha_E = E/H = 5$ V/cm.Oe, i.e., in a 1 cm thick sample a 1 Oe field will produce a 5 V signal. In Table 1, room temperature $\alpha_E$ values reported for some bulk composites are given. The measured values are quite small compared to theoretical predictions.

*Table 1: ME voltage coefficient for bulk composites [Ref. 2,5 and references therein]*

| composite | $\alpha_E$ (V/cm.Oe) | remarks |
|---|---|---|
| $BaTiO_3$-$CoFe_2O_4$ | 0.13 | unidirectional solidification |
|  | 0.05 | sintered |
|  | 0.08 | excess $TiO_2$ |
|  | 0.002-0.01 | hetrocoagulation, ballmilled |
| $BaTiO_3$-$(Ni,Co)Fe_2O_4$ | 0.025 | sintered |

The enormous reduction in the measured $\alpha_E$ is not surprising for the following reasons.
- The bulk composites are extensively microcracked leading to poor mechanical coupling.
- Composites grown by unidirectional solidification of eutectic melts developed microcracks due to thermal expansion mismatch between the two phases.
- The piezoelectric phase is susceptible to electrical shunting by low resistivity spinel ferrites. In addition, possible presence of $Fe^{2+}$ in the ferrite phase will increase the leakage current through the sample and compensate piezoelectrically generated charges.
- Porosity, any impurity, or undesired phases will also decrease $\alpha_E$.
- The connectivity assumed in the theory for the constituent phases cannot be controlled in a bulk composite.

With proper choice for magnetostrictive and piezoelectric phases and a multilayer composite geometry, it might be possible to achieve an $\alpha$-value much higher than observed in bulk composites. Our work on multilayer ME composites is considered next.

## 2. Multilayer Composites

A multilayer (ML) configuration for ME composites has several advantages over bulk samples. In particular, the inherently low piezoelectric constant for bulk composites due to leakage currents through low resistivity ferrites can easily be overcome in the ML structures. Moreover control over mechanical and electrical connectivities between the two phases can easily be realized. Thus it is possible to synthesize the exact structures for which theories predict high ME coefficients. We describe here ML structures in general and results of recent experimental studies.

A composite with alternate layers of magnetostrictive (M) and piezoelectric (P) phases, as shown in Fig. 1, is called a 2-2 composite since the two phases have mechanical connectivity only in the plane of the layers. In spite of a series mechanical connectivity, the P- layers can have either series or parallel electrical connectivity. A series electrical connectivity in which the resistance of P- and M- layers are in series is desirable for high resistance for the ML structure, but the overall piezoelectric constant for the ML will be diminished due to the ferrite layers.

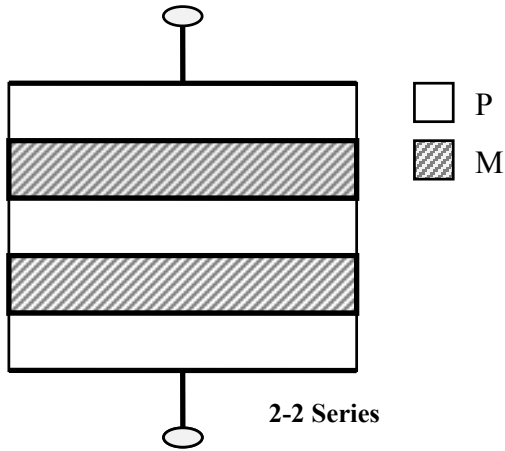

Fig. 1: Multilayer composites of piezoelectric (P-) and magnetostrictive (M-) phases with 2-2 mechanical connectivity, i.e., the two phases are in contact in two dimensions in the layer plane. The electrical connectivity between M- and P- layers is series. [after Ref. 5]

It is, however, possible to enhance the piezoelectric effects by completely shorting electrically the MS layers, so that the P- layers are essentially connected in series [4]. The resulting ML structure, as shown in Fig. 1(a), is a 'mixed' composite with 2-2 mechanical and 3-0 electrical connectivity. Finally, the poling direction in the PE layer and the bias magnetic field direction in MS layers can be varied independently to achieve several different configurations for the ML structures.

Harshe et al., synthesized ML structures consisting of $CoFe_2O_4$-PZT or $BaTiO_3$. Cobalt ferrite has cubic spinel structure and has one of the highest magnetostriction reported for any oxide. PZT was chosen for its high piezoelectric effect and $\varepsilon_r$ values. Doctor-blade techniques were used to make 25-100 μm thick tapes of both phases and platinum electrodes were deposited or painted. Tapes were laminated, sintered at 1500 K, polished, and poled in an electric field. The structures fabricated in this study included (i) composites with mechanical 2-2 and electrical 3-0 connectivities as in Fig. 1(a), (ii) 2-2 composites with parallel electrical connectivity similar to the scheme as in Fig. 1(b), and (iii) other structures with a variety of boundary conditions.

Important observations reported in these studies by Harshe, et al., are as follows [5]. (i) $CoFe_2O_4$-$BaTiO_3$ structures did not show either piezoelectric or magnetoelectric effects. (ii) 2-2 composites with parallel connectivity, similar to Fig. 1(b), did not sinter well. (iii) The highest ME coefficient $\alpha_E$ was measured for $CoFe_2O_4$-PZT with 2-2 mechanical and 3-0 electrical connectivity. The open circuit voltage developed across the sample was measured for a bias magnetic field of $H_o$ = 0-10 kOe and an ac magnetic field of H = 10-15 Oe. The measured $\alpha_E$ of 0.075 V/cm.Oe was a factor of 2-7 smaller that the expected value. (iv) A further reduction in $\alpha_E$ was observed for composites of the same phases, but for other mechanical boundary conditions.

The authors attributed the low $\alpha_E$-values to impurities or undesired phases formed during high temperature sintering and the presence of electrodes which makes the structure a trilayer ML configuration leading to poor mechanical coupling between M-P layers. The formation of highly anisotropic magnetic phases such as barium ferrite at the interface could lead to degradation of ME effects. Additional problems arise due to the use of a foreign electrode (platinum) at the interface. (i) The electrode makes it a 3-phase multilayer structure and leads to poor mechanical coupling between the two oxide layers. (ii) Platinum with a much higher thermal expansion coefficient than the oxides will result in microcracks at the interface during sample processing. (iii) Measurement conditions for ME effects might correspond to the inelastic region of stress-strain characteristics for Pt leading to a reduction in ME coefficients. The degradation of ME effects because of the metal electrodes is also evident from a recent study on multilayers of nickel ferrite-PZT in which a maximum $\alpha_E$ of 136 mV/cm.Oe, a factor of 2 larger than in structures with Pt electrodes, was reported [6]. In summary the use of appropriate M- and P- phases and the elimination of foreign electrodes are critically important for obtaining large ME effects in the ML structures.

The multilayer structure proposed by Harshe, et al., is a major development in terms of synthesis of composites with desirable ME properties. The most significant advantage is the high degree of poling and the consequent increase in piezoelectric and ME effects that could be achieved in the structure. There are additional advantages, namely, the choice of electrical connectivity and orientations for E-field (poling) and bias field $H_o$ (magnetostriction). *But the choice of the constituent phases was not appropriate for accomplishing strong ME coupling.*

### 3. Giant ME Effect in Ferrite-PZT Multilayers

During the past three years we have made considerable progress toward accomplishing theoretically predicted giant ME voltage coefficients in layered systems. Systems studied so far include ferrite-lead zirconate titanate (PZT) and manganite-PZT. We synthesized thick film structures by sintering 10-200 micron tapes obtained by doctor blade techniques [7]. We reported a giant ME effect in bilayers and multilayers of nickel ferrite and PZT. The magnetoelectric voltage coefficient $\alpha_E$ rangeed from 460 mV/cm Oe in bilayers to 1500 mV/cm Oe for multilayers. The transverse effect was an order of magnitude stronger than longitudinal $\alpha_E$. Data on the dependence of $\alpha_E$ on volume fraction of the two phases and bias magnetic field were in excellent agreement with a theoretical model for a bilayer [8].

The observation of strong magnetoelectric (ME) coupling was also reported in cobalt zinc ferrite, $Co_{1-x}Zn_xFe_2O_4$ (CZFO) (x=0-0.6) or nickel zinc ferrite $Ni_{1-x}Zn_xFe_2O_4$ (NZFO) (x=0-0.5) with PZT. The ME voltage coefficient $\alpha_E$ was measured for transverse and longitudinal field orientations for frequencies 10–1000 Hz. A substantial enhancement in $\alpha_E$ was observed with the substitution of Zn. The largest increase, by about 500% was observed in CZFO-PZT and the smallest increase of 60% for NZFO-PZT. As the Zn concentration was increased, $\alpha_E$ increased and showed a maximum for x = 0.2-0.4, depending on the ferrite. The Zn-assisted enhancement in ME coefficient was discussed in terms of Joule magnetostriction, initial permeability and magneto-mechanical coupling for the ferrites [9-11].

Another system of interest was manganite-PZT. The observation of strong magnetoelectric effects was reported in layered ferromagnetic manganite-PZT. Studies were made on $La_{0.7}Sr_{0.3}MnO_3$ (LSMO)-PZT and $La_{0.7}Ca_{0.3}MnO_3$ (LCMO)-PZT. The ME coupling was stronger in LSMO-PZT than in LCMO-PZT, and was weaker in multilayers compared to bilayers [12]. Details of these studies are provided next.

### 3.1 Synthesis of Multilayers

Multilayer samples were synthesized using thick films of ferrites or manganites and PZT obtained by tape casting [7]. The ferrite powder necessary for tape casting was prepared by the standard ceramic techniques that involved mixing the oxides or carbonates of the constituent metals, followed by presintering and final sintering. A ballmill was used to grind the powder to submicron size. For PZT films, we used commercially available powder [13]. The fabrication of thick films contained the following main steps: a) preparation of cast of constituent oxides; b) deposition of 10-40 μm thick films tapes by doctor blade techniques; and c) lamination and sintering of composites. Ferrite or PZT powders were mixed with a solvent (ethyl alcohol), plasticizer (butyl benzyl phthalate), and binder (polyvinyl butyral) in a ball mill for 24 hrs. The slurries were cast into 10-40 μm tapes on silicon coated mylar sheets using a tape caster. The films were dried in air for 24 hrs, removed from the mylar substrate and arranged to obtain the desired structure. They were then laminated under high pressure (3000 psi) and high temperature (400 K), and sintered at 1200-1475 K. Multilayers consisted of (n+1) layers of ferrites/manganites and n layers of PZT (n = 5-30). The samples were characterized in terms of structural, electric, magnetic and magnetoelectric parameters.

The surface morphology and cross-section of the samples were examined with a high-resolution (x1000) metallurgical microscope. Samples contained fine grains and some open pores. The porosity ranged from 5 to 10% depending on the processing temperature. In particular, CZFO-PZT samples showed high porosity. The cross-section studies showed well-bonded structure with uniform thickness for ferrite and PZT. Structural characterization was carried out on sintered multilayers and powdered samples using an x-ray diffractometer. Data for sintered composites did not show any epitaxial or textured nature for the films. Data for powdered samples showed two sets of well-defined peaks; the first set of narrow peaks corresponded to the magnetic phase (NZFO or CZFO), while the second set of relatively broad peaks was identified with the piezoelectric phase (PZT). Main peaks of both sets were of nearly equal intensity. The key inference from x-ray diffraction studies is that no detectable new (impurity) phases are formed because of anticipated diffusion at the interface.

The structural parameters for bulk ferrites, and ferrite and PZT films in the layered samples were calculated from the x-ray data and are shown in Fig.1. The figure also shows the Zn concentration dependence of the full-width-at-half-maximum (FWHM) for the most intense ferrite and PZT diffraction peaks for the composites. Both bulk and thick films of ferrites have a cubic spinel structure and the lattice constants increases linearly with Zn concentration, in very good agreement with reported values [14]. The narrowness of the peaks and the estimated lattice parameters indicate that the spinel structure is preserved and that the ferrite layers are free of interface strain. But the situation is different for PZT. Bulk PZT is found to be tetragonal with a = 0.4062(5) nm, c = 0.4105(6) nm and a FWHM of $0.3^0$ for the (110) peak The data in Fig.2 shows a general broadening of the PZT peaks in the composites. Although the c-value for PZT is unchanged in the layered samples, the data shows a smaller a-value for both series of multilayers. In NZFO-PZT, the reduction in a- amounts to 1% in the unit cell volume of PZT and it remains independent of Zn. A larger volume reduction, as much as 2.5%, occurs in CZFO-PZT. One also notices a linear increase in a-value with Zn concentration in CZFO-PZT. Thus x-ray data implies a strained PZT in the heterostructures. The strain is rather large in CZFO-PZT compared to NZFO-PZT.

The diffraction pattern for powder samples sintered at 1250-1350 K contained reflections from LSMO (LCMO) and PZT phases. For higher $T_s$, impurity phases were present. These observations are in agreement with low temperature magnetization data that showed a reduction in the saturation magnetization for samples sintered at $T_s$ = 1350-1500 K and ferromagnetic resonance data that revealed line broadening. Thus samples sintered at high temperatures contained impurity phases that resulted from diffusion across the manganite-PZT interface.

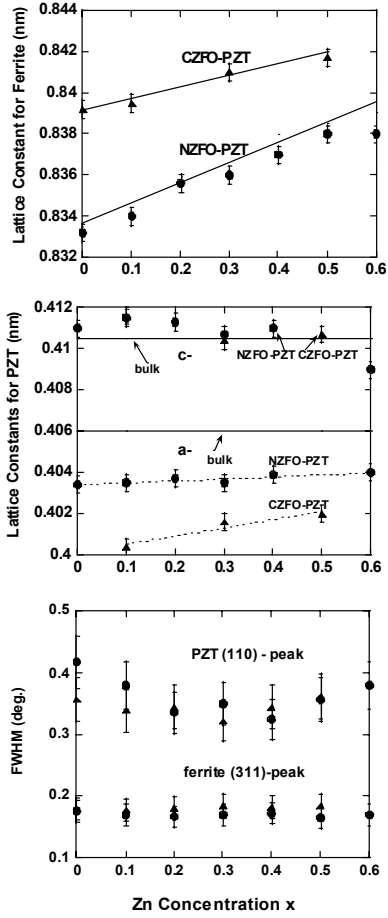

*Fig.2: Lattice constants and the full-width-at-half-maximum (FWHM) as a function of Zn substitution x for ferrites and lead zirconate titanate (PZT) in multilayer composites. The solid lines are lattice constants for bulk nickel zinc ferrite, $Ni_{1-x}Zn_xFe_2O_4$ (NZFO), cobalt zinc ferrite, $Co_{1-x}Zn_xFe_2O_4$ (CZFO) and PZT. The triangles and circles in the upper part are a-values for the cubic spinels CZFO and NZFO, respectively, in the multilayers. The middle part shows the tetragonal a- and c-values for PZT in NZFO-PZT (circles) and CZFO-PZT (triangles). The lower part shows the FWHM-values for the (110) diffraction peak for PZT and (311) peak for CZFO (triangles) and NZFO (circles) in the composites.*

Magnetic characterization included magnetization with a Faraday balance and a vibrating sample magnetometer, ferromagnetic resonance at x-band and magnetostriction with a strain gage. The saturation magnetization $M_s$ were found to agree with bulk values and increased initially with Zn substitution for x=0-0.4, followed by a decrease for higher Zn substitution [9]. For CZFO-PZT samples, there was a significant dependence of magnetization on magnetic field orientation due to the expected high magnetic anisotropy. Investigation of ferromagnetic resonance (FMR) in composites was carried out using a spectrometer operating at 9.8 GHz. The derivative of the microwave absorption line as a function of the magnetic field was registered for the static field H parallel and perpendicular to the sample plane. Absorption spectra for CZFO-PZT had a very wide and irregular shape, while NZFO-PZT had rather narrow and symmetrical lines. This correlates with observed hysteresis and coercivity in magnetization for CZFO-PZT and their absence for NZFO-PZT. Data on ferromagnetic resonance fields were used to estimate the anisotropy field $H_A$. For NZFO-PZT, it varied from 400 Oe for x=0 to 50 Oe for x=0.5 For CZFO-PZT, $H_A$ decreased from 3.7 kOe to 400 Oe as x was increased from 0 to 0.5. Magnetostriction was measured with the standard strain gage technique and detailed data are provided in Section IV. Measurements of electrical resistance R and capacitance C were carried out to probe the quality of the composites. The R and C were smaller than expected values due to either higher than expected conductivity of PZT films or the presence of "shorts" in the PZT films. The ferroelectric-to-paraelectric transition temperature of 600 K agreed with bulk value [15].

Samples were polished; electrical contacts were made with silver paint, and poled. The poling procedure involved heating the sample to 420 K and the application of an electric field E of 20 kV/cm. As the sample was cooled to 300 K, E was increased progressively to 50 kV/cm over 30 min. The piezoelectric coupling coefficient was measured with a $d_{33}$-meter. The parameter of importance for the multilayers is the magnetoelectric voltage coefficient $\alpha_E$. Magnetoelectric measurements are usually performed under two distinctly different conditions: the induced magnetization is measured for an applied electric field or the induced polarization is obtained for an applied magnetic field. We measured the electric field produced by an alternating magnetic field applied to a biased composite. The samples were placed in a shielded 3-terminal sample holder and placed between the pole pieces of an electromagnet that was used to apply a bias magnetic field H. The required ac field of $\delta H$=1 Oe at 10-1000 Hz parallel to H was generated with a pair of Helmholtz coils. The resulting ac electric field $\delta E$ perpendicular to the sample plane was estimated from the voltage $\delta V$ measured with a lock-in-amplifier. The ME voltage coefficient $\alpha_E = \delta V/t\delta H$ where t is the *effective thickness* of PZT. The measurements are done for two different field orientations. With the sample plane represented by (*1,2*), the transverse coefficient $\alpha_{E,31}$ is measured for the magnetic fields H and $\delta H$ along direction-*1* (parallel to the sample plane) and perpendicular to $\delta E$ (direction-*3*). The longitudinal coefficient $\alpha_{E,33}$ is measured for all the fields perpendicular to the sample plane. Magnetoelectric characterization is carried out as a function of frequency and bias magnetic field H.

### 3.2 Nickel Ferrite – PZT

Figure 3 shows the static magnetic field dependence of the transverse ME coefficient $\alpha_{E,31}$ for a two layer composite of NFO-PZT, each layer with a thickness of 200 μm [8]. The data at room temperature are for a frequency of 1 kHz and *for unit*

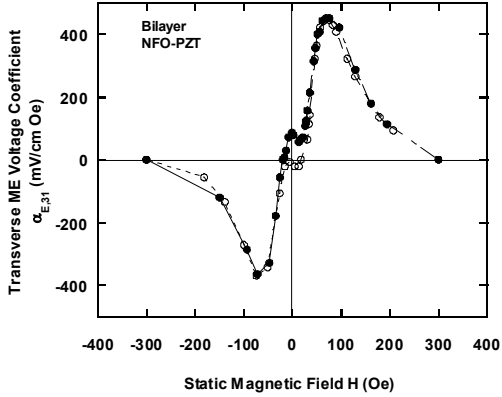

*Fig. 3: Transverse magnetoelectric (ME) voltage coefficient $\alpha_{E,31} = \delta E_3/\delta H_1$ at room temperature as a function of static magnetic field H for a two-layer structure consisting of 200 μm films of nickel ferrite (NFO) and lead zirconate titanate (PZT). The field H and a 1-kHz ac magnetic field $\delta H_1$ are applied parallel to each other and parallel to the sample plane and the induced electric field $\delta E_3$ is measured perpendicular to the sample plane. The open (filled) circles are the data points for increasing (decreasing) H. The lines are guides to the eye.*

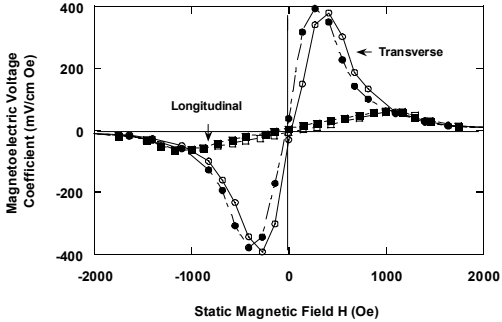

*Fig. 4: Static field dependence of room temperature transverse and longitudinal ME voltage coefficients, $\alpha_{E,31}$ and $\alpha_{E,33}$ respectively, for a multilayer composite consisting of alternate layers of NFO and PZT. The sample contained 15 layers of NFO and 14 layers of PZT. The thickness of each layer is 14 μm. For longitudinal ME effects the fields H, $\delta H$ (1 kHz) and $\delta E$ are parallel to each other and perpendicular to the sample plane. The open (filled) circles are the data points for increasing (decreasing) H. The lines are guides to the eye. The inset shows frequency and temperature dependence of $\alpha_{E,31}$ measured at 1 kHz.*

*thickness of the piezoelectric phase.* As H is increased from zero, $\alpha_{E,31}$ increases, reaches a maximum value of 460 mV/cm.Oe at 70 Oe and then drops rapidly to zero above 300 Oe. There is no evidence for hysteresis in Fig. 1 except at fields close to zero. We observed a phase difference of 180 degrees between the induced voltages for +H and -H. As discussed later, the magnitude and the field dependence in Fig.

3 are related to the slope of λ vs. H characteristics and can be understood in terms of pseudo-piezomagnetic effects in nickel ferrite.

Similar field dependence of both the transverse and longitudinal ME voltage coefficients, $\alpha_{E,31}$ and $\alpha_{E,33,}$ respectively, are shown in Fig. 4 for a multilayer composite. The data at room temperature for a 1 kHz ac field are for a sample consisting of alternate layers of NFO (15) and PZT (14), each layer with a thickness of 14 μm so that the effective thickness of PZT is the same as for the bilayer. Comparison of data in Figs.3 and 4 indicate the following. In multilayers, (i) ME effects are observed over a wider field range, (ii) the field for maximum $\alpha_{E,31}$ is shifted to higher fields and (iii) the peak value of $\alpha_{E,31}$ is 15% smaller than for the bilayer. One observes in Fig.4 a noticeable hysteresis in the field dependence of $\alpha_{E,31}$ and $\alpha_{E,33.}$

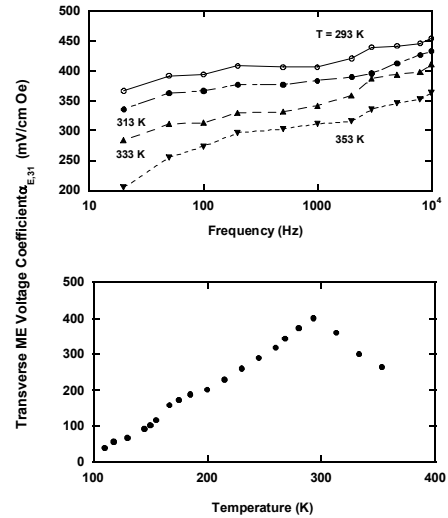

*Fig.5: Frequency and temperature dependence of the transverse coefficient $\alpha_{E,31}$ for the multilayer with n = 14. The lines are guides to the eye. Values of $\alpha_{E,31}$ are for the bias field H corresponding to maximum value in the ME effect. The temperature dependence is for a frequency of 100 Hz.*

The variation of the multilayer $\alpha_{E,31}$ with frequency and temperature are shown in Fig.5. Upon increasing the frequency from 20 Hz to 10 kHz, an overall increase of 25% occurs in $\alpha_{E,31}$. The ME coefficient is found to be maximum at room temperature and it decreases with increasing temperature. The variations are most likely due to the frequency and temperature dependence of material parameters for the constituent phases.

### 3.3: Nickel Zinc Ferrite-PZT

Similar ME studies were performed on nickel zinc ferrite-PZT samples with x = 0-0.5 [9,10]. Figure 6 shows representative data on the H-dependence of $\alpha_E$ for NZFO-PZT samples with x = 0-0.4. The data were obtained on samples with n=10-15 at room temperature for a frequency of 100 Hz. For NFO-PZT (x=0), $\alpha_{E,31}$ vs H shows the expected resonance-like character with a maximum centered at H = 400 Oe. When Zn is substituted for Ni, we notice an increase in the peak value of $\alpha_{E,31}$ for low x-values. As x is increased, a down-shift is observed in H-value corresponding to the peak in $\alpha_{E,31}$. The magnetic field range for strong ME effects decreases with increasing Zn content. Data on the longitudinal coupling in Fig.6 shows the following important departures from the transverse case. (i) The coupling strength does not show any dependence on x for low Zn substitution. (ii) As x is increased, an up-shift is observed for the H-value corresponding to peak $\alpha_{E,33}$. (iii) The ME coupling is realized over a wide field range [9,10].

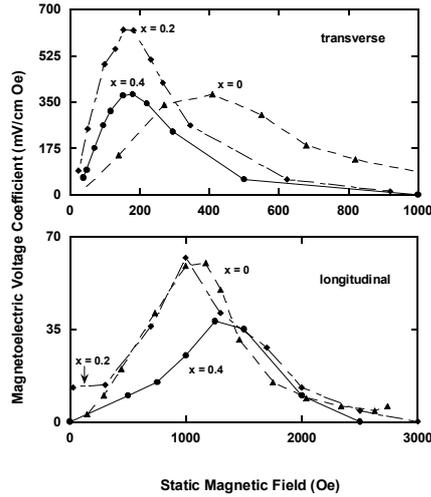

*Fig.6: ME voltage coefficients versus H data for multilayer samples of $Ni_{1-x}Zn_xFe_2O_4$ (NZFO) - PZT.*

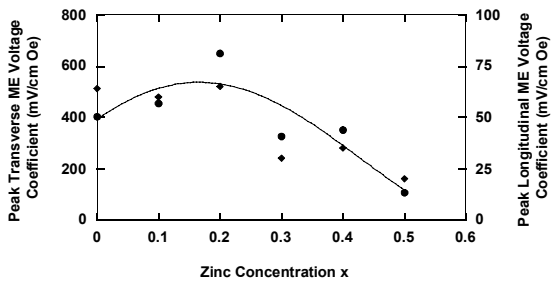

*Fig.7: Zinc concentration dependence of maximum transverse and longitudinal ME coefficients in NZFO – PZT layered samples. The line is guide to the eye.*

Similar $\alpha_E$ vs H profiles were obtained for other x values [9,10]. The variations of peak values of $\alpha_E$ with x are plotted in Fig.7. Average values of $\alpha_E$ and the spread are plotted. The data reveal a 60% increase in the transverse ME voltage coefficient as x is increased from 0 to 0.2, followed by a reduction in $\alpha_{E,31}$ for higher x. The longitudinal coefficient also shows a similar character. The $\alpha_E$ values are significantly higher than reported values in past studies on bulk or layered samples. The coupling coefficient must be compared with 80 mV/cm Oe for NFO-barium titanate, 115 mV/cm Oe for NFO-PZT bulk composites, and 300-400 mV/cm Oe for NFO-PZT bilayers and multilayers [2,8].

### 3.4 Cobalt Zinc Ferrite-PZT

Cobalt ferrite is ideal for use as the ferromagnetic phase in the composite due to high magnetostriction. Studies were performed on $Co_{1-x}Zn_xFe_2O_4$ (CZFO) (x=0-0.6)-PZT [9,11]. Figure 8 shows representative data on the H dependence of $\alpha_{E,31}$ and $\alpha_{E,33}$ for a sample in which 40% Co is replaced by Zn. The data at room temperature and 100 Hz are for a sample with n = 10. As the bias field is increased from zero, $\alpha_E$ increases rapidly to a peak value. With further increase in H, the ME coefficients drop to a minimum or zero value. When H is reversed, we observed (i) a 180 deg. phase difference relative to the ME voltage for +H and (ii) a small decrease in the peak value for $\alpha_E$ compared to the value for +H. There was no noticeable hysteresis or remenance in the $\alpha_E$ vs. H behavior. Similar $\alpha_E$ vs H data were obtained for samples with x-values varying from 0 to 0.6. Both $\alpha_{E,31}$ and $\alpha_{E,33}$ were measured. Figure 9 shows the room temperature variation of $\alpha_{E,31}$ with the bias magnetic field for x = 0-0.4. Data on the longitudinal coupling are not shown since the coupling was quite weak for all x-values except for 0.4. As x is increased one notices (i) an increase in the rate at which $\alpha_{E,31}$ varies with H at low static magnetic fields, (ii) the peak in $\alpha_{E,31}$ occurs at progressively decreasing H, and (iii) there is a general increase in the peak value of $\alpha_{E,31}$. In Fig.10 the variation of maximum $\alpha_E$ with x is shown for both transverse and longitudinal cases. The peak coefficients were measured for several samples, both multilayers and bilayers, and the figure shows the average value and the spread that amounts to ±10-20%. As the Zn substitution is increased, one observes a sharp increase in $\alpha_{E,31}$, from 50 mV/cm Oe for x = 0 to 280 mV/cm Oe for x = 0.4. Further increase in x is accompanied by a substantial reduction in $\alpha_{E,31}$. A similar character is evident for the longitudinal coupling parameter.

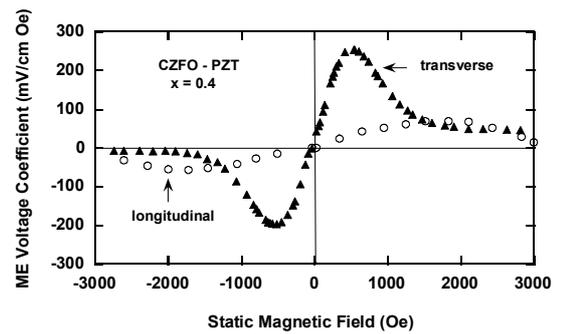

*Fig.8: Magnetoelectric (ME) voltage coefficient $\alpha_E = \delta E/\delta H$ versus bias magnetic field H for a multilayer of $Co_{1-x}Zn_xFe_2O_4$ (CZFO) (x=0.4)–lead zirconate titanate (PZT). The sample contained 11 layers of CZFO and 10 layers of PZT with a thickness of 18 μm. The data at room temperature and 100 Hz are for transverse (out-of-plane δE perpendicular to in-plane δH) and longitudinal (out-of-plane δE and δH) field orientations. There is 180 deg. phase difference between voltages for +H and –H.*

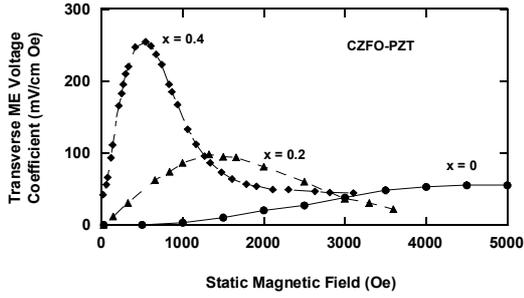

*Fig.9: Transverse ME voltage coefficient $\alpha_{E,31}$ as a function of H at room temperature and 100 Hz for multilayers of CZFO – PZT with x = 0, 0.2 and 0.4.*

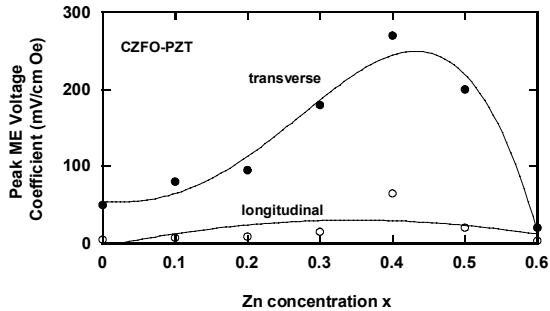

*Fig.10: Variation of peak transverse and longitudinal ME voltage coefficients with zinc concentration x in layered CZFO – PZT.*

Past attempts on cobalt ferrite based ME composites included bulk samples with barium titanate or PZT and multilayers with PZT [5]. Bulk samples with barium titanate showed evidence for ME coupling, but reported $\alpha_E$ were very small. Bulk composites with PZT showed a similar very weak ME effect, but layered samples showed a maximum $\alpha_E$ of 75 mV/cm Oe [5]. It is obvious from data in Figs.8-10 that Zn substitution in cobalt ferrite is a key ingredient for strong ME coupling in multilayers. We attribute the efficient field conversion properties to modification of magnetic parameters due to Zn substitution.

### 3.5 Lithium Zinc Ferrite-PZT

Finally, we consider studies on lithium zinc ferrite-PZT composites [16]. Samples with n=10 and 15, a layer thickness of 15 micron and ferrites $Li_{0.5-x/2}Zn_xFe_{2.5-x/2}O_4$ for x=0-0.4 were synthesized. Figure 11 shows data on H-variation of the transverse ME voltage coefficient for x = 0-0.3. The data are for a frequency of 100 Hz at room temperature. Important observations are as follows. (i) Data show features similar to the other two systems. (ii) A factor of five increase in the peak $\alpha_{E,31}$ value is evident when x is increased from 0 to 0.3. (iii) An up-shift in the H-value corresponding the peak $\alpha_{E,31}$ is seen as x is increased from 0. Recall that for CZFO-PZT and NZFO-PZT samples, a down-shift in H for maximum $\alpha_{E,31}$ is observed for increasing x. (iv) The H-interval for strong ME effects is essentially independent of x. The inset in Fig.10 shows data on peak values of $\alpha_{E,31}$ vs x. One notices a rapid increase in the ME voltage coefficient with x for x = 0- 0.3, followed by a sharp decrease for x=0.4 [16].

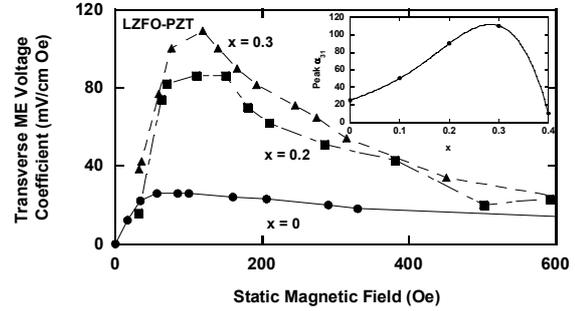

*Fig. 11: Transverse voltage coefficient versus H profiles as in Figs. 1 and 4, but for multilayer samples of $Li_{0.5-x/2}Zn_xFe_{2.5-x/2}O_4$ (LZFO) - PZT. The inset shows peak value of $\alpha_{E,31}$ vs x.*

### 3.6 Discussion

We now discuss the results for ferrites-PZT. We developed a theoretical model for ME coupling in the bilayers and multilayers. According to the model, $\alpha_E$ is directly proportional to the product of piezomagnetic and piezoelectric coefficients, and is dependent on the volume for both phases. The order of magnitude difference in transverse ans longitudinal coefficients essentially arises due to the directional dependence of the magnetostriction $\lambda$ and piezomagnetic coefficients.

Next we address an important question that arises from the study. What is the cause of giant ME effects in NFO-PZT and NZFO-PZT? The large $\alpha_{E,31}$ is in part due to inherent advantages in ML geometry: efficient poling and the total absence of leakage currents. But other systems such as CFO-PZT also have the same advantages. In ferrites, domains are spontaneously deformed in the magnetization direction and the Joule magnetostriction is caused by domain wall motion and domain rotation in the presence of H. Since ME effects involve dynamic magnetoelastic coupling, key requirements for the ferrite phase are unimpeded domain motion and a large $\lambda$. A soft, high initial permeability (low coercivity) and high $\lambda$-ferrite, such as NFO used here, is the main ingredient for strong ME effects. In magnetically hard cobalt ferrite, however, one has the disadvantage of a large coercive field that limits domain rotation. Since the ME effect originates at the interface, it is important to consider the influence of growth-induced stress and its effect on magnetic anisotropy and the dynamics of domain motion. The interface coupling is also influenced by a variety of other factors such as defects, inhomogeneities, and grain boundaries that pin the domain and limit wall motion and rotation. In the absence of detailed studies on NFO-PZT and other ferrite-PZT interfaces, we could only speculate on possible reasons for a unique and favorable interface bonding in the present system.

Important factors that affect the interface properties during the high temperature processing are the thermal expansion coefficients (2 ppm for PZT vs 10 ppm for most ferrites), thermal conductivity (an order of magnitude higher in CFO compared to PZT or NFO), and the sintering temperature. Differential thermal expansion and thermal conductivity could result in built-in strain and interface microcracks. The sintering temperature of 1425-1500 K is much closer to the melting temperature of cobalt ferrite (1840 K) than NFO (2020 K) and it is quite likely that the highly reactive lead compound (PZT) could easily form both structural and chemical inhomogeneities at the interface with cobalt ferrite. So it is reasonable to conclude that the giant ME effect in NFO-PZT is most likely due to an interface free of growth induced stress or defects and a favorable domain dynamics. Investigations on the microscopic nature of the NFO-PZT interface with techniques such as high-resolution x-ray diffraction, electron microscopy, and magnetic force microcopy are critically important for an understanding of the current observations.

### 4. Manganite-PZT

Lanthanum manganites with divalent substitutions have attracted considerable interest in recent years due to double exchange mediated ferromagnetism, metallic conductivity, and giant magnetoresistance. The manganites are potential candidates for ME composites because of (i) high magnetostriction and (ii) metallic conductivity that eliminates the need for a foreign electrode at the P-M interface. We reported strong ME effects and unique magnetic field dependence in composites of $La_{0.7}Sr_{0.3}MnO_3$ (LSMO)–PZT and $La_{0.7}Ca_{0.3}MnO_3$ (LCMO)- PZT [12]. Samples were prepared by sintering tapecast films. X-ray diffraction was performed for structural characterization. The diffraction pattern for powder samples sintered at 1250-1350 K contained reflections from LSMO (LCMO) and PZT phases. For higher $T_s$, impurity phases were present. These observations are in agreement with low temperature magnetization data that showed a reduction in the saturation magnetization for samples sintered at $T_s$ = 1350-1500 K and ferromagnetic resonance data that revealed line broadening. Thus samples sintered at high temperatures contained impurity phases that resulted from diffusion across the manganite-PZT interface. The data reported here are for the sample sintered at 1325 K, free of detectable amount of impurities and with the best magnetic parameters for LSMO (LCMO).

*LSMO-PZT layered composites*: The best ME parameters are obtained at low temperatures for the bilayer sample [12]. Figure 12 shows the H dependence of the transverse coefficient $\alpha_{E,31}$ for a bilayer (n = 2) of $La_{0.7}Sr_{0.3}MnO_3$–PZT. The data are for temperatures 120 K and 300 K and for a frequency of 100 Hz for the ac magnetic field. Consider first the data at 120 K. As H is increased from –500 Oe, $\alpha_{E,31}$ increases in magnitude and peaks at –35 Oe. One observes a large remanence at H= 0. With further increase in H, $\alpha_{E,31}$ goes through zero value at the coercive field $H_c$ = 35 Oe and is accompanied by a 180 deg. phase shift in the ME voltage. A peak in $\alpha_{E,31}$ is evident at 140 Oe and it decreases gradually to zero with further increase in H. One essentially observes similar features for decreasing H, from 500 Oe to –500 Oe, but the peak is down shifted in H and the peak value is higher than for increasing fields. The most significant observations are the hysteresis and remanence in $\alpha_E$ vs H. The loop is asymmetric with a squareness-ratio of well over 90%. When the sample temperature is increased to 300 K, one essentially observes similar features as for 120 K, but with the following departures: (i) a pronounced asymmetry in the hysteresis loop and a reduction in $H_c$, and (ii) an overall reduction in the magnitude of $\alpha_{E,31}$.

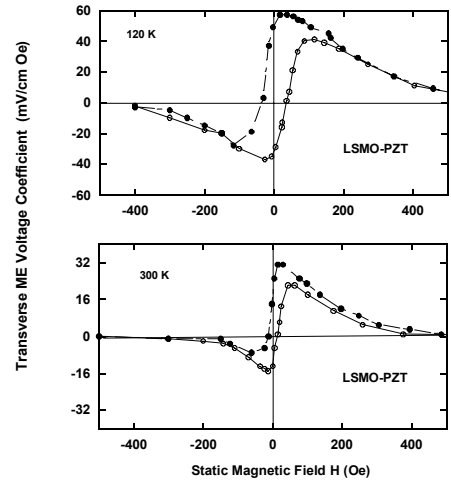

*Fig.12: Transverse magnetoelectric (ME) voltage coefficient at 120 K and 300 K as a function of static magnetic field H for a two-layer structure consisting of 200 μm films of lanthanum strontium manganite $La_{0.7}Sr_{0.3}MnO_3$ (LSMO) and lead zirconate titanate (PZT). The open (filled) circles are the data points for increasing (decreasing) H. The lines are guides to the eye.*

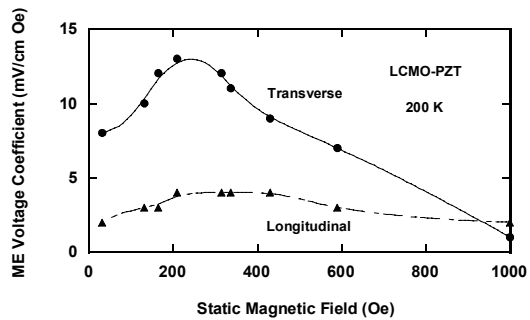

*Fig.13: Static magnetic field dependence of transverse and longitudinal ME voltage coefficients measured at 200 K and 100 Hz for a bilayer of $La_{0.7}Ca_{0.3}MnO_3$ (LCMO)-PZT. The layer thickness was 200 micron each for LCMO and PZT. The lines are guides to the eye.*

*LCMO-PZT composites*: We synthesized similar layered composites of $La_{0.7}Ca_{0.3}MnO_3$ (LCMO) – PZT and investigated the ME properties. Bilayers and multilayers with n = 4, 6 were prepared. The variation of ME voltage coefficients with the bias magnetic field is shown in Fig.13 for a bilayer with a layer thickness of 200 μm each for LCMO and PZT. The data were obtained for a sample temperature of 200 K. The transverse coefficient peaks around 200 Oe. The longitudinal coupling is quite weak, a factor of 3-6 smaller than $\alpha_{E,31}$.

It is clear from the results presented here that there is a strong ME coupling in manganites-PZT layered composites. The most significant observations are (i) a relatively large transverse ME coefficient compared to the longitudinal coupling, (ii) a stronger ME coupling in LSMO-PZT than in LCMO-PZT and (iii) weakening of ME effects in multilayers. As discussed earlier for ferrite-PZT, the H dependence of $\alpha_E$ essentially tracks the slope of $\lambda$ vs.H. We estimated both the transverse and longitudinal ME coefficients and were an order magnitude higher than measured values.

We now discuss possible causes for the discrepancy between theory and data. Since the field conversion is an interface phenomenon, it is necessary to focus on the details of the LSMO-PZT interface for an understanding of the disagreement between theory and data. It is important to consider the influence of growth-induced strain at the interface and its effect on ferromagnetic order, magnetic anisotropy and the dynamics of domain motion. Detailed studies on the nature of LSMO-PZT interface is lacking at present. One must also consider the effects of interface defects, inhomogeneities, and grain boundaries that pin the domains and limit wall motion and rotation. Magnetic force microscopy studies on LCMO show evidence for domain pinning by surface defects. As we mentioned earlier, degradation of magnetic parameters in samples sintered above 1350 K indicates diffusion of metal ions across the LSMO-PZT boundary and the formation of chemical inhomogeneities and defects. Even though the samples in the present study (sintered at 1325 K) showed expected values for magnetic parameters, presence of impurities due to diffusion at the interface cannot be totally ruled out [12].

*Acknowledgments:* The work was supported by NSF grants (DMR-0072144, DMR-0302254).